\documentclass[12pt]{article}

\usepackage{graphicx}
\usepackage{amsmath}
\usepackage{amssymb}
\allowdisplaybreaks

\begin{document}

\baselineskip=15pt plus 0.2pt minus 0.1pt

\begin{titlepage}
\title{
\hfill\parbox{4cm}
{\normalsize KEK-TH-1075\\
{\tt hep-th/0603068}}\\
\vspace{1cm}
{\bf Twist Field as Three String Interaction Vertex\\
in Light Cone String Field Theory}
}
\author{
Isao {\sc Kishimoto}${}^{1}${}\thanks
{{\tt ikishimo@post.kek.jp}},~ 
Sanefumi {\sc Moriyama}${}^{2}${}\thanks
{{\tt moriyama@math.nagoya-u.ac.jp}},~
Shunsuke {\sc Teraguchi}${}^{3}${}\thanks
{{\tt teraguch@phys.ntu.edu.tw}}
\\[12pt]
{${}^{1}$
\it High Energy Accelerator Research Organization (KEK),}\\
{\it Tsukuba 305-0801, Japan}\\[6pt]
{${}^{2}$
\it Graduate School of Mathematics, Nagoya University,}\\
{\it Nagoya 464-8602, Japan}\\[6pt]
{${}^{3}$
\it National Center for Theoretical Sciences, Taiwan}\\
}
\date{\normalsize March, 2006}
\maketitle
\thispagestyle{empty}

\begin{abstract}
\normalsize
It has been suggested that matrix string theory and light-cone string
field theory are closely related.
In this paper, we investigate the relation between the twist field,
which represents string interactions in matrix string theory, and the
three-string interaction vertex in light-cone string field theory
carefully.
We find that the three-string interaction vertex can reproduce some of
the most important OPEs satisfied by the twist field.
\end{abstract}

\end{titlepage}

\section{Introduction}
Retrospecting recent progress in understanding various interesting
effects in string theory, we are led to the desire of constructing a
complete off-shell formulation of string theory.
As will be explained below, at present there are two formulations for
light-cone quantization of type IIB closed superstring theory.
However, neither of them is considered to be complete.

One of the formulations is the light-cone superstring field theory
(LCSFT) \cite{GS,GSB} constructed from supersymmetry algebra.
The starting point is the Green-Schwarz action for free strings.
We can construct the Hamiltonian and the supercharges satisfying the
supersymmetry algebra out of it.
The interaction terms are added to these charges by requiring that the 
total charges satisfy the supersymmetry algebra perturbatively.
The first order interaction term is given as
\begin{align}
|H_1\rangle_{123}&=Z^i\bar Z^jv^{ij}(\Lambda)|V\rangle_{123}\,,
\label{SFTH1}\\
|Q_1^{\dot\alpha}\rangle_{123}
&=\bar Z^is^{i\dot\alpha}(\Lambda)|V\rangle_{123}\,,\\
|\tilde Q_1^{\dot\alpha}\rangle_{123}
&=Z^i\tilde s^{i\dot\alpha}(\Lambda)|V\rangle_{123}\,.
\end{align}
Here $|V\rangle_{123}$ is the three-string interaction vertex
constructed by the overlapping condition and $Z^i$ $(\bar Z^i)$ is the
holomorphic (anti-holomorphic) part of the bosonic momentum at the
interaction point, whose divergence is regularized as follows:
\begin{align}
\Bigl(P^i+\frac{1}{2\pi\alpha}X^{i\prime}\Bigr)(\sigma)|V\rangle_{123}
\sim\frac{1}{\sqrt{\sigma-\sigma_{\rm I}}}Z^i|V\rangle_{123}\,,
\label{Zi}
\end{align}
with $\alpha=p^+$ and $\sigma_I$ being the interaction point.
$\Lambda$ is the regularization of the fermionic momentum at the
interaction point and $v^{ij}(\Lambda)$, $s^{i\dot\alpha}(\Lambda)$
and $\tilde s^{i\dot\alpha}(\Lambda)$ are known but intricate
functions of $\Lambda$.
The program of constructing the interaction terms is successful at the
first order, though it is too complicated to proceed to higher
orders.

The other formulation is matrix string theory (MST) \cite{MSSBS,DVV},
which stems from the Matrix formulation of light-cone quantization of
M-theory \cite{BFSS} and takes the form of $(1+1)$-dimensional super
Yang-Mills theory.
To relate MST to the perturbative string, we first note that the
Yang-Mills coupling $g_{{\rm YM}}$ is related to the string coupling
$g_{{\rm s}}$ and the string length $\sqrt{\alpha'}$ by
$g_{{\rm YM}}^{-1}=g_{{\rm s}}\sqrt{\alpha'}$.
Hence, the free string limit corresponds to the IR limit and the first 
order interaction term to the least irrelevant operator.
{}From the requirement of the dimension counting and the locality of
the interaction, we expect that the first order interaction term is
written as dimension three operator constructed essentially out of the
twist field.
The interaction term of MST is proposed to be
\cite{DVV}
\begin{align}
H_1=\sum_{m,n}\int d\sigma
\bigl(\tau^i\Sigma^i\bar\tau^j\bar\Sigma^j\bigr)_{m,n}\,,
\label{MSTH1}
\end{align}
where $\tau^i$ is the excited twist field defined as
\begin{align}
\partial X^i(z)\cdot\sigma(0)\sim\frac{1}{\sqrt{z}}\tau^i(0)\,,
\label{taui}
\end{align}
with $\sigma(z,\bar z)$ being the ${\mathbb Z}_2$ twist field and
$\Sigma^i(z)$ being the spin field for the Green-Schwarz fermions.
The indices $m$ and $n$ of the twist fields denote the string bits
where the ``exchange'' interaction takes place.
These indices have to be summed over in calculating the string
amplitude.

The expression \eqref{MSTH1} in MST seems somewhat formal compared
with that in LCSFT \eqref{SFTH1}, though it is more promising to go
beyond the first order in MST than in LCSFT \cite{DM}.
Hopefully we can obtain some information in LCSFT from MST.
For this purpose, we would like to relate LCSFT to MST carefully.
We can easily find a close analogy between \eqref{SFTH1} and
\eqref{MSTH1} and between \eqref{Zi} and \eqref{taui}, if we regard
$\sigma(z,\bar z)$ as $|V\rangle_{123}$ and $\tau^i(z,\bar z)$ as
$Z^i|V\rangle_{123}$.
Following this analogy between LCSFT and MST, two supercharges of MST
were written down explicitly in \cite{M}.
These arguments of supercharges are consistent with the pioneering but
primitive argument in \cite{DVV} and with the relation between LCSFT
and MST proposed in \cite{DM}.

In this paper, we would like to proceed further to investigate the
relation between the twist field $\sigma(z,\bar z)$ and the
three-string interaction vertex $|V\rangle_{123}$ scrupulously.
In particular, in addition to the defining OPE of the excited twist
field \eqref{taui}, we would like to realize the OPE of two twist
fields  \cite{DFMSBR,OZ} (for each dimension)
\begin{align}
\sigma(z,\bar z)\cdot\sigma(0)
\sim\frac{1}{|z|^{1/4}(\ln |z|)^{1/2}}\,,
\label{twosigma}
\end{align}
in terms of the three-string interaction vertex $|V\rangle_{123}$.
To realize the OPE \eqref{twosigma}, we identify the interaction point
$\sigma_I$ of $|V\rangle_{123}$ with the insertion point $z$ of the
twist field $\sigma(z,\bar z)$.
We multiply two string interaction vertices with a short intermediate
time $T$ to see whether the effective interaction vertex reproduces
the reflector (which corresponds to the identity operator in CFT) with
the suitable singularity.

A natural question arises here. 
In LCSFT the three-string interaction vertex $|V\rangle_{123}$ is
always accompanied by the level-matching projection
\begin{align}
{\cal P}_r=\oint{d\theta\over 2\pi}
e^{i\theta(L_0^{(r)}-\bar{L}_0^{(r)})}\,,
\end{align}
on each string $r=1,2,3$.
We would like to see which one corresponds to the twist field; the
interaction vertex with projections,
${\cal P}_1{\cal P}_2{\cal P}_3|V\rangle_{123}$, or the vertex without
them, $|V\rangle_{123}$.
Our answer to this question is as follows.
To calculate the amplitude in LCSFT, we need to integrate over the
intermediate string length ($\alpha=p^+$) and perform the
level-matching projection at each string.
The level-matching projection is equivalent to integrating over the
diagrams by shifting the interaction point by an angle.
These two integrations are combined into a simple summation over
string bits $m$ and $n$ in \eqref{MSTH1}.
Since the twist field by itself is the expression before the
summations, the corresponding interaction vertex in LCSFT should not
contain any summations.
Hence, the interaction vertex corresponding to the twist field is the
one $|V\rangle_{123}$ without the level-matching projection and the
intermediate string length integration.

There are two ways to realize \eqref{twosigma} because the interaction
vertices can be connected in two different ways.
One of them is the four-point tree diagram connecting the long string
of two interaction vertices (fig.~\ref{fig:4string}) and the other
is the two-point 1-loop diagram connecting two short strings
(fig.~\ref{fig:2string}).
We shall evaluate these two diagrams in the next section to see that
both of the results are proportional to the reflector with the same
singularity as that in \eqref{twosigma}.

Note that it is desirable to perform all the computations of the above
diagrams in the superstring theory, if we want to relate LCSFT to
MST.
However, here we shall utilize the bosonic string theory for
simplicity \cite{Rey}.
It should not be too difficult to generalize our computation to the
supersymmetric case.
Also note that the separation of the two interaction points in the
above diagrams is in the worldsheet time direction, while we separate
two insertion points in the space direction in \eqref{Zi}.
Since OPE does not depend on in which direction one operator
approaches the other, we assume that the results do not depend on the
directions.

The content of this paper is as follows.
In the next section, we shall first recapitulate some necessary
ingredients of LCSFT.
The first subsection is devoted to the computation of the tree diagram
and in the second and third subsections we compute the 1-loop
diagram.
Finally, we conclude with some further directions.
A short review of Neumann coefficients is given in appendix A.
A somewhat related result about free field realization of boundary
changing operators is given in appendix B.

\section{LCSFT computation}
In this section, we would like to evaluate the two diagrams mentioned
in the introduction to see the correspondence between the twist field
$\sigma(z,\bar z)$ and the interaction vertex $|V\rangle_{123}$.
For this purpose, let us briefly review the closed LCSFT here.
Three-string interaction vertex for LCSFT with $\alpha=p^+$ fixed is
given as
\begin{align}
|V(1_{\alpha_1},2_{\alpha_2},3_{\alpha_3})\rangle
=[\mu(\alpha_1,\alpha_2,\alpha_3)]^2\int\delta(1,2,3)
e^{E(1,2,3)+\bar{E}(1,2,3)}|p_1\rangle_1|p_2\rangle_2|p_3\rangle_3\,,
\label{3string}
\end{align}
where $\alpha_1+\alpha_2+\alpha_3=0$ and
\begin{align}
&\mu(\alpha_1,\alpha_2,\alpha_3)
=\exp\biggl(-\tau_0\sum_{r=1}^3{1\over \alpha_r}\biggr)\,,
\quad\tau_0=\sum_{r=1}^3\alpha_r\log |\alpha_r|\,,\\
&\int\delta(1,2,3)=\int 
{d^{d-2}p_1\over (2\pi)^{d-2}}
{d^{d-2}p_2\over (2\pi)^{d-2}}
{d^{d-2}p_3\over (2\pi)^{d-2}}
(2\pi)^{d-2}\delta^{d-2}(p_1+p_2+p_3)\,,\\
&E(1,2,3)={1\over 2}\sum_{r,s=1}^3\sum_{m,n\ge 1}N^{r,s}_{mn}
a_m^{i(r)\dagger}a_n^{i(s)\dagger}
+\sum_{r=1}^3\sum_{n\ge 1}N^r_na^{i(r)\dagger}_n{\mathbb P}^i_{123}
-{\tau_0\over 2\alpha_1\alpha_2\alpha_3}{\mathbb P}^2_{123}\,,
\end{align}
with ${\mathbb P}^i_{123}=\alpha_1p_2^i-\alpha_2p_1^i$
$(i=1,\cdots,d-2)$.
We define the normalized left-moving oscillators
$a_n^i=\alpha_n^i/\sqrt{n}$, $a_n^{i\dagger}=\alpha_{-n}^i/\sqrt{n}$
for $n\ge 1$ satisfying
$[a_n^i,a_m^{j\dagger}]=\delta_{n,m}\delta^{i,j}$ and
$a_n^i|p\rangle=0$.
$\bar a_n^i$ is the right-moving cousin and $\bar E(1,2,3)$ is
defined by replacing $a_n^i$ by $\bar a_n^i$ in $E(1,2,3)$.

For later convenience, let us note that $E(1,2,3)$ can be recast into
the following form
\begin{align}
E(1,2,3)&=\frac{1}{2}\boldsymbol{a}^{(3)\dagger{\rm T}}N^{3,3}
\boldsymbol{a}^{(3)\dagger}
+\boldsymbol{a}^{(3)\dagger{\rm T}}
N^{3,12}\boldsymbol{a}^{(12)\dagger}
+\frac{1}{2}\boldsymbol{a}^{(12)\dagger{\rm T}}N^{12,12}
\boldsymbol{a}^{(12)\dagger}\nonumber\\
&\quad+\bigl(\boldsymbol{a}^{(3)\dagger{\rm T}}\boldsymbol{N}^{3}
+\boldsymbol{a}^{(12)\dagger{\rm T}}\boldsymbol{N}^{12}\bigr)
{\mathbb P}_{123}
-\frac{\tau_0}{2\alpha_1\alpha_2\alpha_3}{\mathbb P}^2_{123}\,,
\end{align}
if we adopt the matrix notation for the indices of infinite
oscillation modes
\begin{align}
\bigl(\boldsymbol{a}^{(r)}\bigr)_m=a^{(r)}_m\,,\quad
\bigl(\boldsymbol{a}^{(r)\dagger}\bigr)_m=a^{(r)\dagger}_m\,,\quad
\bigl(\boldsymbol{N}^{r}\bigr)_m=N^{r}_m\,,\quad
\bigl(N^{r,s}\bigr)_{mn}=N^{r,s}_{mn}\,,
\end{align}
and define
\begin{align}
&\boldsymbol{a}^{(12)\dagger}=\begin{pmatrix}
\boldsymbol{a}^{(1)\dagger}\\\boldsymbol{a}^{(2)\dagger}
\end{pmatrix}\,,\quad
\boldsymbol{N}^{12}=\begin{pmatrix}
\boldsymbol{N}^{1}\\\boldsymbol{N}^{2}\end{pmatrix}\,,\nonumber\\
&N^{3,12}=\begin{pmatrix}N^{3,1}&N^{3,2}\end{pmatrix}\,,\quad
N^{12,3}=\begin{pmatrix}N^{1,3}\\N^{2,3}\end{pmatrix}\,,\quad
N^{12,12}=\begin{pmatrix}
N^{1,1}&N^{1,2}\\N^{2,1}&N^{2,2}
\end{pmatrix}\,.
\end{align}
In terms of the matrix notation, the Neumann coefficients satisfy
among others\footnote{One may wonder why $N^{12,3}$ is invertible
because it does not look like a square matrix.
To clarify this point, let us regularize the size of the
infinitely-dimensional Neumann matrices by truncating it at a finite
level.
Due to the expression of the mode expansion \eqref{P}, the UV cutoff
of the worldsheet $\Delta\sigma_r$ of string $r$ is related to the
truncation level $L_r$ by $L_r\Delta\sigma_r/|\alpha_r|\sim 1$.
If we fix the UV cutoff of the worldsheet to a constant and choose
string 3 to be the long string, $|\alpha_3|=|\alpha_1|+|\alpha_2|$,
the truncation levels of three strings have to be related by
$L_3=L_1+L_2$.
In this sense $N^{12,3}$ is a square matrix.}
\begin{align}
&N^{3,12}N^{12,3}+N^{3,3}N^{3,3}=1\,,\quad
N^{3,12}N^{12,12}+N^{3,3}N^{3,12}=0\,,\quad
N^{12,12}N^{12,12}+N^{12,3}N^{3,12}=1\,,\nonumber\\
&\qquad\qquad\qquad
N^{3,12}\boldsymbol{N}^{12}+N^{3,3}\boldsymbol{N}^{3}
=-\boldsymbol{N}^{3}\,,\quad
N^{12,12}\boldsymbol{N}^{12}+N^{12,3}\boldsymbol{N}^{3}
=-\boldsymbol{N}^{12}\,,\nonumber\\
&\qquad\qquad\qquad\qquad\qquad\qquad
\boldsymbol{N}^{3{\rm T}}\bigl(N^{12,3}\bigr)^{-1}\boldsymbol{N}^{12}
=-\frac{\tau_0}{\alpha_1\alpha_2\alpha_3}\,,
\label{NN}
\end{align}
which play important roles in our later computation.
More details about the Neumann coefficients can be found in appendix
A.

The reflector for closed LCSFT is given as
\begin{align}
\langle R(1,2)|=\int\delta(1,2)
\,{}_1\langle p_1|{}_2\langle p_2|
e^{-(\boldsymbol{a}^{(1){\rm T}}\boldsymbol{a}^{(2)}
+\bar{\boldsymbol{a}}^{(1){\rm T}}\bar{\boldsymbol{a}}^{(2)})}\,,
\end{align}
with 
\begin{align}
\int\delta(1,2)=\int
{d^{d-2}p_1 \over (2\pi)^{d-2}}
{d^{d-2}p_2 \over (2\pi)^{d-2}}
(2\pi)^{d-2}\delta^{d-2}(p_1+p_2)\,,
\end{align}
and $\langle p|$ defined as 
$\langle p|p'\rangle=(2\pi)^{d-2}\delta^{d-2}(p-p')$,
$\langle p|a^{\dagger}_n=0$ and $\langle p|\bar{a}^{\dagger}_n=0$.

We shall utilize the following formula in our later calculation.
\begin{align}
&\langle 0|\exp\biggl(\frac{1}{2}\boldsymbol{a}^{\rm T}
M\boldsymbol{a}
+\boldsymbol{a}^{\rm T}\boldsymbol{k}\biggr)
\exp\biggl(\frac{1}{2}\boldsymbol{a}^{\dagger{\rm T}}
N\boldsymbol{a}^\dagger
+\boldsymbol{a}^{\dagger{\rm T}}\boldsymbol{l}\biggr)
|0\rangle\nonumber\\
&=[\det(1-MN)]^{-1/2}
\exp\biggl(\boldsymbol{l}^{\rm T}\frac{1}{1-MN}\boldsymbol{k}
+\frac{1}{2}\boldsymbol{k}^{\rm T}N\frac{1}{1-MN}\boldsymbol{k}
+\frac{1}{2}\boldsymbol{l}^{\rm T}\frac{1}{1-MN}M\boldsymbol{l}
\biggl)\,.
\label{product}
\end{align}

\subsection{Tree diagram}
\begin{figure}[htbp]
\begin{center}
\scalebox{0.5}[0.5]{\includegraphics{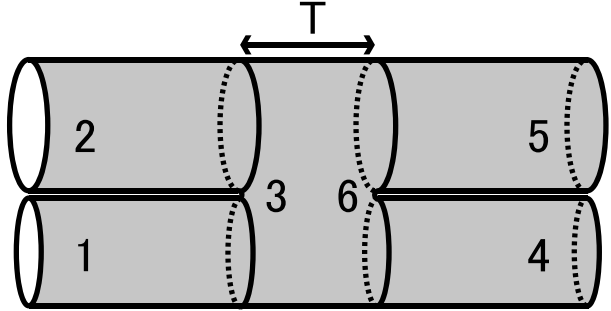}}
\end{center}
\caption{Four-string tree diagram in the $\rho$-plane.}
\label{fig:4string}
\end{figure}

We now turn to the evaluation of the four-point tree diagram with two
incoming strings 1 (with length $\alpha_1(>0)$) and 2 (with length
$\alpha_2(>0)$), joining and splitting again into two outgoing strings
4 and 5 of the same length as 1 and 2, respectively. 
(See fig.~\ref{fig:4string}.)
Note that since only the two twist fields which exchange the same
string bits enjoy the OPE \eqref{twosigma}, we have to choose
$\alpha_4=-\alpha_1$ and $\alpha_5=-\alpha_2$ so that the exchange
interactions take place at the same string coordinate.
The effective four-string interaction vertex of this diagram is given
as ($\alpha'=2$)
\begin{align}
\label{eq:tree}
|A(1,2,4,5)\rangle=\langle R(3,6)|e^{-{T\over |\alpha_3|}
(L_0^{(3)}+\bar{L}_0^{(3)})}
|V(1_{\alpha_1},2_{\alpha_2},3_{\alpha_3})\rangle
|V(4_{-\alpha_1},5_{-\alpha_2},6_{-\alpha_3})\rangle\,,
\end{align}
with $L_0=p^ip^i/2+\sum_{n\ge 1}na_n^{i\dagger}a_n^i-1$ and
$\bar{L}_0=p^ip^i/2+\sum_{n\ge 1}n\bar a_n^{i\dagger}\bar a_n^i-1$.
Note that under the simultaneous change of signs
$\alpha_1\to -\alpha_1\,,\alpha_2\to -\alpha_2\,,
\alpha_3\to -\alpha_3\,,$ $N^{r,s}$ is even while $\boldsymbol{N}^{r}$
is odd.
Hence the Neumann coefficients in
$|V(4_{-\alpha_1},5_{-\alpha_2},6_{-\alpha_3})\rangle$ can be written
in terms of those of
$|V(1_{\alpha_1},2_{\alpha_2},3_{\alpha_3})\rangle$.
{}From the OPE \eqref{twosigma} we expect that $|A(1,2,4,5)\rangle$ is
proportional to the reflector $|R(1,4)\rangle|R(2,5)\rangle$ with the
coefficient divergent as
$\bigl(T^{-1/4}(\ln T)^{-1/2}\bigr)^{24}=\bigl(T(\ln T)^2\bigr)^{-6}$
if we take the limit $T\to +0$.
Here we identify the coordinate $|z|$ of the twist field
$\sigma(z,\bar z)$ with the intermediate propagation time $T$ up to a
numerical factor when we are interested only in the short distance
behavior of two operators.
Though generally $z$ may be related to $T$ in a complicated way, the
relation is approximately linear at a short distance.

In order to perform our calculation simply, let us first rewrite the
proper time expression of the propagator
$e^{-{T\over |\alpha_3|}(L_0^{(3)}+\bar{L}_0^{(3)})}$
into $e^{-{T\over 2|\alpha_3|}
(L_0^{(3)}+\bar{L}_0^{(3)}+L_0^{(6)}+\bar{L}_0^{(6)})}$.
By applying the formula \eqref{product} with
\begin{align}
&\boldsymbol{a}=\boldsymbol{a}^{(36)}\,,\quad
M=\begin{pmatrix}0&-1\\-1&0\end{pmatrix}\,,\quad
N=\begin{pmatrix}N^{3,3}_{T/2}&0\\0&N^{3,3}_{T/2}\end{pmatrix}\,,
\nonumber\\
&\boldsymbol{k}=\boldsymbol{0}\,,\quad
\boldsymbol{l}=\begin{pmatrix}N^{3,12}_{T/2}&0\\
0&N^{3,12}_{T/2}\end{pmatrix}
\begin{pmatrix}\boldsymbol{a}^{(12)\dagger}\\
\boldsymbol{a}^{(45)\dagger}\end{pmatrix}
+\begin{pmatrix}\boldsymbol{N}^{3}_{T/2}{\mathbb P}_{123}\\
-\boldsymbol{N}^{3}_{T/2}{\mathbb P}_{456}\end{pmatrix}\,,
\end{align}
with $N^{3,3}_{T}=e^{-{T\over |\alpha_3|}C}N^{3,3}
e^{-{T\over |\alpha_3|}C}$,
$N^{3,12}_{T}=e^{-{T\over |\alpha_3|}C}N^{3,12}$ and
$\boldsymbol{N}^3_{T}
=e^{-{T\over |\alpha_3|}C}\boldsymbol{N}^3$, we can compute
$|A(1,2,4,5)\rangle$ without difficulty:
\begin{align}
|A(1,2,4,5)\rangle
=A_T\int\delta(1,2,4,5)
e^{F_T(1,2,4,5)}|p_1\rangle_1|p_2\rangle_2|p_4\rangle_4|p_5\rangle_5\,,
\label{eq:tree_T}
\end{align}
with
\begin{align}
&A_T=\left|[\mu(\alpha_1,\alpha_2,\alpha_3)]^2{\det}^{-{d-2\over 2}}
(1-N^{3,3}_{T/2}N^{3,3}_{T/2})\right|^2\,,\\
&\int\delta(1,2,4,5)=\int
{d^{d-2}p_1\over (2\pi)^{d-2}}
{d^{d-2}p_2\over (2\pi)^{d-2}}
{d^{d-2}p_4\over (2\pi)^{d-2}}
{d^{d-2}p_5\over (2\pi)^{d-2}}
(2\pi)^{d-2}\delta^{d-2}(p_1+p_2+p_4+p_5)\,.
\end{align}
The exponent $F_T(1,2,4,5)$ takes a complicated expression.
However, in the limit $T\to +0$ we can evaluate it formally with the
use of \eqref{NN} and 
${\mathbb P}_{123}-{\mathbb P}_{456}=\alpha_3(p_1+p_4)$
which holds because of the momentum conservation
$(2\pi)^{d-2}\delta^{d-2}(p_3+p_6)$.
The formal result of it is as follows.
\begin{align}
&\lim_{T\rightarrow +0}F_T(1,2,4,5)=
-(\boldsymbol{a}^{(12)\dagger{\rm T}}\boldsymbol{a}^{(45)\dagger}
+\bar{\boldsymbol{a}}^{(12)\dagger{\rm T}}
\bar{\boldsymbol{a}}^{(45)\dagger})\nonumber\\
&\qquad-(p_1+p_4)\alpha_3
\boldsymbol{N}^{3{\rm T}}\bigl(N^{12,3}\bigr)^{-1}
(\boldsymbol{a}^{(12)\dagger}
+\boldsymbol{a}^{(45)\dagger}
+\bar{\boldsymbol{a}}^{(12)\dagger}
+\bar{\boldsymbol{a}}^{(45)\dagger})\nonumber\\
&\qquad-(p_1+p_4)^2\alpha_3^2\boldsymbol{N}^{3{\rm T}}
\bigl(1-(N^{3,3})^2\bigr)^{-1}
\boldsymbol{N}^{3}\,.
\label{T0}
\end{align}
Note that the first term gives the nonzero modes of the
reflector. 
This is the first sign that our expectation works. 
Also the last term will give the zero mode part of the reflector
$\delta^{d-2}(p_1+p_4)$ if the quantity $b_0=\alpha_3^2
\boldsymbol{N}^{3{\rm T}}\bigl(1-(N^{3,3})^2\bigr)^{-1}
\boldsymbol{N}^{3}$ is divergent.
Actually this seems to be true from numerical analysis.
We can fit as $b_0\simeq(1\sim 3)\log L+({\rm constant})$, where $L$
is the size of Neumann matrices which we used in our numerical
computation.

To regularize it properly, let us retrieve the intermediate time $T$
in our calculation.
We find that, instead of the divergent quantity $b_0$, we have
$b_T=\alpha_3^2\boldsymbol{N}^{3{\rm T}}_{T/2}
\bigl(1-(N^{3,3}_{T/2})^2\bigr)^{-1}
\boldsymbol{N}^3_{T/2}$ which, according to (C.18), (C.20) and (C.21)
in \cite{HIKKO2}, is identified to be
\begin{align}
b_T=\alpha_3^2\boldsymbol{N}^{3{\rm T}}_T
\bigl(1-N^{3,3}N^{3,3}_T\bigr)^{-1}
\boldsymbol{N}^3=-\log(1-Z_5)\,.
\label{regularization}
\end{align}
Here we have mapped the worldsheet in the ``light-cone'' type
$\rho$-plane into the whole complex $z$-plane by the Mandelstam map
\begin{align}
\rho(z)=\alpha_1\bigl(\log(z-Z_1)-\log(z-Z_4)\bigr)
+\alpha_2\bigl(\log(z-Z_2)-\log(z-Z_5)\bigr)\,,
\label{mandelstam}
\end{align}
and fixed the gauge by choosing $Z_1=\infty, Z_2=1, Z_4=0$ and
$0<Z_5<1$.
Note that, without the insertion of the level-matching projections,
the moduli parameter, $Z_5$, runs only along the real axis.
To see the behavior of \eqref{regularization} in the limit $T\to +0$,
all we have to do is to relate $T$ to $Z_5$ as in Chapter 11 of
\cite{GSW}.
For this purpose, we note that $T$ can be regarded as the difference
of two stationary points $z_\pm$ in the $\rho$-plane:
\begin{align}
T=\rho(z_+)-\rho(z_-)\,,\quad
\frac{d\rho}{dz}\Bigr|_{z=z_\pm}\!\!\!=0\,.
\label{stationary}
\end{align}
Since the limit $T\to +0$ corresponds to the $t$-channel limit, $Z_5$
should approach $Z_2=1$ in this limit.
After an explicit calculation, we find
\begin{align}
\frac{T}{|\alpha_3|}\sim 
4\sqrt{\frac{\alpha_1\alpha_2}{\alpha_3^2}}\sqrt{1-Z_5}\,,
\end{align}
and hence $b_T\sim 2\log(|\alpha_3|/T)$.
This is consistent with the numerical result if we identify the
regularization parameter by ${|\alpha_3|}/T\sim L$.
Therefore, the contribution of the exponential of the last term in
\eqref{T0} is
\begin{align}
e^{-b_T(p_1+p_4)^2}
\sim\biggl[\frac{\pi}{2\log(|\alpha_3|/T)}\biggr]^{\frac{d-2}{2}}
\delta^{d-2}(p_1+p_4)\,.
\end{align}
This implies that 
\begin{align}
(2\pi)^{d-2}\delta^{d-2}(p_1+p_2+p_4+p_5)e^{-b_T(p_1+p_4)^2}
\propto(2\pi)^{d-2}\delta^{d-2}(p_1+p_4)
(2\pi)^{d-2}\delta^{d-2}(p_2+p_5)\,,
\end{align}
which gives the zero mode part of the reflector.

The determinant factor $A_T$ in \eqref{eq:tree_T} is already evaluated
as ($d=26$)
\begin{align}
A_T\sim 2^{10}\biggl|\frac{\alpha_1\alpha_2}{\alpha_3^2}\biggr|^2
\left[T\over|\alpha_3|\right]^{-6},
\end{align}
in \cite{CG, HIKKO2}. (See also appendix B of \cite{KMW2}.)
Combining all the contributions, we find that in the limit $T\to +0$,
\begin{align}
|A(1,2,4,5)\rangle\sim 2^{-26}\pi^{-12}
\biggl[\frac{T}{|\alpha_{123}|}
\biggl(\log\frac{T}{|\alpha_{123}|}\biggr)^2\,\biggr]^{-6}
|R(1,4)\rangle|R(2,5)\rangle\,,
\label{treeresult}
\end{align}
with $\alpha_{123}=(\alpha_1\alpha_2\alpha_3)^{1/3}$.
This is consistent with our expectation.

\subsection{1-loop diagram}
\begin{figure}[htbp]
\begin{center}
\scalebox{0.5}[0.5]{\includegraphics{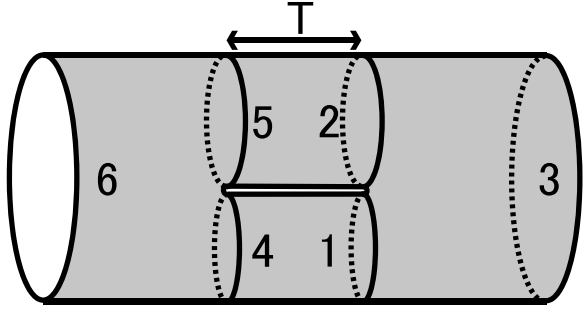}}
\end{center}
\caption{Two-string 1-loop diagram in the $\rho$-plane.}
\label{fig:2string}
\end{figure}

In the previous subsection, we have computed one realization of the
OPE \eqref{twosigma}.
Here we would like to proceed to the other realization via the 1-loop
diagram: the incoming string 6 splits into two short strings and join
again into the outgoing string 3. 
(See fig.~\ref{fig:2string}.)
For this purpose, let us calculate $(\alpha_1,\alpha_2>0)$
\begin{align}
\label{eq:1-loop}
|B(3,6)\rangle=\langle R(2,5)|\langle R(1,4)|
e^{-{T\over \alpha_1}(L_0^{(1)}+\bar{L}_0^{(1)})
-{T\over \alpha_2}(L_0^{(2)}+\bar{L}_0^{(2)})}
|V(1_{\alpha_1},2_{\alpha_2},3_{\alpha_3})\rangle
|V(4_{-\alpha_1},5_{-\alpha_2},6_{-\alpha_3})\rangle\,.
\end{align}

The calculation is parallel to the previous case of the tree diagram.
Using \eqref{product} we obtain
\begin{align}
|B(3,6)\rangle
=B_T\int{d^{d-2}p_1 \over(2\pi)^{d-2}}\int\delta(3,6)
\,e^{F_{T}(3,6,p_1)}|p_3\rangle_3|p_6\rangle_6\,,
\label{eq:1-loop_T}
\end{align}
with 
\begin{align}
B_T=\left|[\mu(\alpha_1,\alpha_2,\alpha_3)]^2\,{\det}^{-{d-2\over 2}}
\left(1-N^{12,12}N^{12,12}_T\right)\right|^2\,,
\end{align}
and $N^{12,12}_T
={\rm diag}(e^{-{T\over\alpha_1}C},e^{-{T\over\alpha_2}C})N^{12,12}
{\rm diag}(e^{-{T\over\alpha_1}C},e^{-{T\over\alpha_2}C})$.
Again, the exponent $F_T(3,6,p_1)$ can be evaluated in the limit 
$T\to +0$ using various formulas of Neumann coefficients \eqref{NN} as
\begin{align}
\lim_{T\rightarrow +0}F_T(3,6,p_1)
=-(\boldsymbol{a}^{(3)\dagger}\boldsymbol{a}^{(6)\dagger}
+\bar{\boldsymbol{a}}^{(3)\dagger}
\bar{\boldsymbol{a}}^{(6)\dagger})\,.
\label{FT0}
\end{align}
Namely, \eqref{eq:1-loop_T} is proportional to $|R(3,6)\rangle$
including the zero mode sector.
However, the integration of the loop momentum $p_1$ gives a divergent
constant $\delta^{d-2}(0)$ for $T=0$.
Therefore, we need to regularize $|B(3,6)\rangle$ by the intermediate
time $T$ again:
\begin{align}
F_{T}(3,6,p_1)&=F_{T}(3,6,p_1)|_{{\rm osc}}
+c_{T}\biggl[p_1-\frac{\alpha_1}{\alpha_3}p_3\biggr]^2\!
+\boldsymbol{C}_{T}^{\rm T}
(\boldsymbol{a}^{\dagger(3)}-\boldsymbol{a}^{\dagger(6)}
+\bar{\boldsymbol{a}}^{\dagger(3)}-\bar{\boldsymbol{a}}^{\dagger(6)})
\biggl[p_1-\frac{\alpha_1}{\alpha_3}p_3\biggr]\nonumber\\
&~~~~+\biggl[\frac{p_3^2}{\alpha_3}-\frac{2\alpha_3}{\alpha_1\alpha_2}\biggl]T,
\end{align}
with $F_T(3,6,p_1)|_{\rm osc}$ being the oscillator bilinear part of
$F_T(3,6,p_1)$ and $c_T$ and $\boldsymbol{C}_{T}$ being
\begin{align}
c_{T}&=2\alpha_3^2\biggl({T/2-\tau_0\over\alpha_1\alpha_2\alpha_3}
+\boldsymbol{N}^{12{\rm T}}
(1-N^{12,12}_TN^{12,12})^{-1}
(\boldsymbol{N}^{12}_T+N^{12,12}_T\boldsymbol{N}^{12})\biggr)\,,\\
\boldsymbol{C}_{T}
&=\alpha_3\left[\boldsymbol{N}^{3}
+N^{3,12}(1-N^{12,12}_TN^{12,12})^{-1}
(\boldsymbol{N}^{12}_T+N^{12,12}_T\boldsymbol{N}^{12})\right],
\end{align}
where we have used 
${\mathbb P}_{123}={\mathbb P}_{456}=\alpha_3p_1-\alpha_1p_3$ and
defined
$\boldsymbol{N}^{12}_T
={\rm diag}(e^{-{T\over\alpha_1}C},e^{-{T\over\alpha_2}C})
\boldsymbol{N}^{12}$.
Note that our previous result \eqref{FT0} is equivalent to the
following statement.
\begin{align}
&\lim_{T\to +0}F_{T}(3,6,p_1)|_{{\rm osc}}=
-(\boldsymbol{a}^{(3)\dagger}\boldsymbol{a}^{(6)\dagger}
+\bar{\boldsymbol{a}}^{(3)\dagger}
\bar{\boldsymbol{a}}^{(6)\dagger})\,,\quad
\lim_{T\to +0}c_{T}=0\,,\quad
\lim_{T\to +0}\boldsymbol{C}_{T}=\boldsymbol{0}\,.
\end{align}
After we perform the loop momentum $p_1$ integration, the result
$|B(3,6)\rangle$ for $T\ne 0$ becomes
\begin{align}
|B(3,6)\rangle
&=B_T(4\pi c_{T})^{-{d-2\over 2}}
\int\delta(3,6)\,e^{-\frac{1}{4c_T}\left(\boldsymbol{C}_T^{{\rm T}}
(\boldsymbol{a}^{\dagger(3)}-\boldsymbol{a}^{\dagger(6)}
+\bar{\boldsymbol{a}}^{\dagger(3)}-\bar{\boldsymbol{a}}^{\dagger(6)})
\right)^2+(\frac{p_3^2}{\alpha_3}-\frac{2\alpha_3}{\alpha_1\alpha_2})T}\nonumber\\
&~~~~\times
e^{F_{T}(3,6,p_1)|_{{\rm osc}}}
|p_3\rangle_3|p_6\rangle_6\,.
\end{align}

If we can further prove that ($n,m\ge 1$)
\begin{align}
\lim_{T\to +0}\frac{(\boldsymbol{C}_{T})_m(\boldsymbol{C}_{T})_n}
{c_T}=0\,,
\label{CC/c}
\end{align}
we have
\begin{align}
|B(3,6)\rangle\sim K_T|R(3,6)\rangle\,,
\label{propreflect}
\end{align}
with $K_T=B_T(4\pi c_{T})^{-{d-2\over 2}}$ for $T\to +0$.
This assumption \eqref{CC/c} seems to be true from our numerical
analysis, though it is still desirable to prove it algebraically.
The numerical analysis strongly suggests that our result is
proportional to the reflector $|R(3,6)\rangle$.
In the next subsection, we would like to turn to the evaluation of the
leading order of $K_T$ for $d=26$, to see the singular behavior of the
OPE \eqref{twosigma}.

\subsection{Evaluation of $K_T$}
It is difficult to calculate $K_T$ in \eqref{propreflect} directly
using the Neumann coefficients.
For the evaluation of $K_T$ let us contract $|B(3,6)\rangle$ with two
tachyon states.
Since the full propagator including the light-cone directions is given
as
\begin{align}
\Delta_r=\frac{1}{-2p^+_rp^-_r+L_0^{(r)}+\bar L_0^{(r)}}
=\int_0^\infty\frac{dT_r}{\alpha_r}
e^{-\frac{T_r}{\alpha_r}(-2p^+_rp^-_r+L_0^{(r)}+\bar L_0^{(r)})}\,,
\end{align}
the total amplitude (without applying the level-matching projections
${\cal P}_1{\cal P}_2$ on string 1 and 2) is given as
\begin{align}
S_{36}&={}_3\langle-k_3|{}_6\langle-k_6|
\langle R^{{\rm LC}}(2,5)|\langle R^{{\rm LC}}(1,4)|
\Delta_1\Delta_2
|V^{{\rm LC}}(1,2,3)\rangle
|V^{{\rm LC}}(4,5,6)\rangle
\nonumber\\
&=(2\pi)^d\delta^d(k_3+k_6)\int d\alpha_1
\int_0^\infty dT_1\int_0^\infty dT_2\delta(T_2-T_1)
\frac{e^{2T_1k_3^-}}{4\pi\alpha_1\alpha_2}K_T\,,
\label{eq:det_eval}
\end{align}
where $T=T_1=T_2$ and the light-cone directions are included in the
tachyon state $\langle k|$, the reflector $\langle R^{{\rm LC}}|$ and
the interaction vertex $|V^{{\rm LC}}\rangle$.
According to \cite{KZ} this light-cone expression can be calculated in
the $\alpha=p^+$ HIKKO string field theory:
\begin{align}
&S_{36}={}_3\langle -k_3|{}_6\langle -k_6|
\langle R^{\,\alpha=p^+}(2,5)|\langle R^{\,\alpha=p^+}(1,4)|
\frac{b_0^{(1)}\bar b_0^{(1)}}
{L_0^{{\rm tot}(1)}+\bar L_0^{{\rm tot}(1)}}
\frac{b_0^{(2)}\bar b_0^{(2)}}
{L_0^{{\rm tot}(2)}+\bar L_0^{{\rm tot}(2)}}\nonumber\\
&\qquad\qquad\times
|V^{\alpha=p^+}(1,2,3)\rangle
|V^{\alpha=p^+}(4,5,6)\rangle\,.
\label{alpha=p+}
\end{align}
Here the reflector $\langle R^{\,\alpha=p^+}|$ and the interaction
vertex $|V^{\alpha=p^+}\rangle$ are those of the $\alpha=p^+$ HIKKO
string field theory and $L_0^{{\rm tot}}$ and $\bar L_0^{{\rm tot}}$
are the total Virasoro operators including the light-cone directions
of the matter part and the ghost part.

\begin{figure}[htbp]
\begin{center}
\scalebox{0.6}[0.6]{\includegraphics{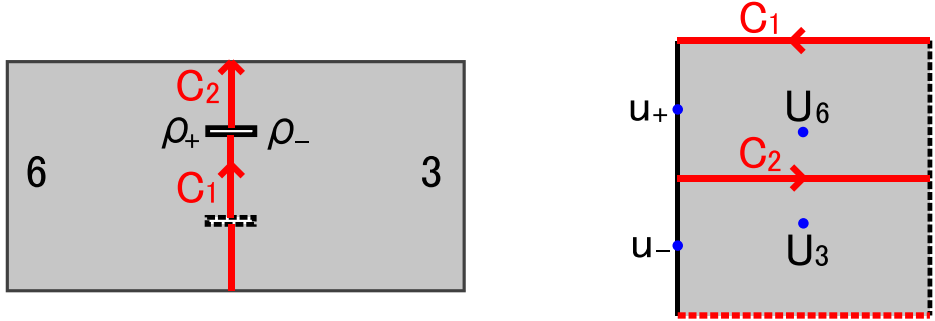}}
\end{center}
\caption{The Mandelstam map from the $\rho$-plane to the $u$-plane.
The left figure is the light-cone $\rho$-plane corresponding to
fig.~\ref{fig:2string} while the right one represents the $u$-plane of 
the torus with periods $1$ and $\tau$.
They are related by $\rho(U_6)=-\infty$, $\rho(U_3)=\infty$,
$\rho(u_+)=\rho_+$ and $\rho(u_-)=\rho_-$.}
\label{fig:rhou}
\end{figure}

We can calculate it with CFT by mapping the light-cone $\rho$-plane
into the torus $u$-plane, where the incoming string 6 and the outgoing
string 3 in the $\rho$-plane are mapped to $U_6$ and $U_3$ in the
$u$-plane respectively.
(See fig.~\ref{fig:rhou}.)
The Mandelstam map $\rho(u)$ \cite{Mandelstam}, which corresponds to
fig.~\ref{fig:rhou}, is given as
\begin{align}
\label{eq:Mandel1loop}
\rho(u)=|\alpha_3|\log
\frac{\vartheta_1(u-U_6|\tau)}{\vartheta_1(u-U_3|\tau)}
-2\pi i\alpha_1u\,.
\end{align}
Here $\tau$ and $U_6-U_3$ are pure imaginary because we do not insert
the level-matching projections ${\cal P}_1{\cal P}_2$ in
\eqref{alpha=p+}.
The moduli parameter $\tau$ of the torus is related to $T$ in
fig.~\ref{fig:2string} by finding the stationary points of the
Mandelstam map as in \eqref{stationary},
\begin{align}
T=\rho(u_-)-\rho(u_+)\,,\quad
\frac{d\rho}{du}\Bigr|_{u=u_\pm}\!\!\!=0\,.
\end{align}
In the degenerating limit $T\to +0$, we have \cite{KM2}:
\begin{align}
\label{eq:degene}
e^{-\frac{i\pi}{\tau}}
\sim\frac{T}{8|\alpha_3|\sin(\pi\alpha_1/|\alpha_3|)}\,.
\end{align}

Using the Mandelstam map \eqref{eq:Mandel1loop} and the proper time
representation of $1/(L_0^{{\rm tot}(r)}+\bar L_0^{{\rm tot}(r)})$,
\eqref{alpha=p+} can be put into the CFT expression on the torus
$u$-plane:
\begin{align}
S_{36}=\int_0^\infty dT_1\int_0^\infty dT_2
\Big\langle(\alpha_1\alpha_2)^{-1}(\alpha_1\alpha_2)^2
b_{T_1}\bar b_{T_1}b_{T_2}\bar b_{T_2}
V(k_3;U_3,\bar U_3)V(k_6;U_6,\bar U_6)\,{\cal C}\Big\rangle_\tau\,.
\label{apCFT}
\end{align}
Note that we have to put the measure $(\alpha_1\alpha_2)^{-1}$ inside
the CFT correlator because the Mandelstam map \eqref{eq:Mandel1loop}
depends on $\alpha_r$.
Here $T_1$ and $T_2$ are the worldsheet proper time of the propagating
strings 1 and 2 while $b_{T_i}$, $\bar b_{T_i}$ and 
$V(k;u,\bar u)$ are defined as
\begin{align}
&b_{T_i}=\int_{C_i}{du\over 2\pi i}{du\over d\rho}b(u)\,,\quad
\bar b_{T_i}
=\int_{\bar C_i}{d\bar u\over 2\pi i}{d\bar u\over d\bar\rho}
\bar b(\bar u)\,,\\
&V(k;u,\bar{u})=
c(u)\bar{c}(\bar{u}):e^{ik_{\mu}X^{\mu}(u,\bar{u})}:\,,
\end{align}
with the integration contours $C_i$ and $\bar C_i$ shown in
fig.~\ref{fig:rhou}.
Note that in the $\alpha=p^+$ HIKKO string field theory, the
propagator $b_0\bar b_0/(L_0^{{\rm tot}}+\bar L_0^{{\rm tot}})$ is
originally defined on the light-cone $\rho$-plane, but the vertex
operator $V(k;u,\bar u)$ comes from that constructed on the unit
disk.
Therefore we have to take the conformal factor into account.
The factor $(\alpha_1\alpha_2)^2$ is from mapping four of $b_0$ in the
$\rho$-plane of each string into the total $\rho$-plane, while the
factor ${\cal C}$ is the conformal factor for the tachyon vertices
mapped from the local unit disk $w_r$ of each string into the torus
$u$-plane:
\begin{align}
{\cal C}=\left|
\left.{du\over dw_3}\right|_{u=U_3}
\left.{du\over dw_6}\right|_{u=U_6}
\right|^{k_3^2-2}\!\!,
\end{align}
where
\begin{align}
\left.{du\over dw_3}\right|_{u=U_3}
\left.{du\over dw_6}\right|_{u=U_6}\!\!\!=
-\biggl({\vartheta_1(U_6-U_3|\tau)
\over\vartheta_1'(0|\tau)}\biggr)^2
e^{2\pi i\frac{\alpha_1}{|\alpha_3|}(U_6-U_3)-\frac{T}{|\alpha_3|}}\,,
\end{align}
with $\vartheta_1'(\nu|\tau)=\partial_\nu\vartheta_1(\nu|\tau)$ and
\begin{align}
\rho=\begin{cases}\alpha_3\log w_3+T/2&\quad{\rm Re}\,\rho>T/2\\
-\alpha_3\log w_6-T/2&\quad{\rm Re}\,\rho<-T/2\end{cases}\,.
\end{align}

Now all we have to do is to evaluate each sector of \eqref{apCFT}.
This was done explicitly in \cite{AKT3} for the open string case.
The ghost sector and the nonzero mode contribution of the matter
sector are exactly the square of the open string case.
The only difference comes from the zero mode contribution of the
matter sector and still it can be evaluated similarly to the open
string case.
The contribution from the ghost part is (See (6.5) in \cite{AKT3}.)
\begin{align}
\Big\langle b_{T_1}\bar{b}_{T_1}b_{T_2}\bar{b}_{T_2}
c(U_3)\bar{c}(\bar{U}_3)c(U_6)\bar{c}(\bar{U}_6)\Big\rangle_\tau
=\left|{R\over 2\pi}{\cal G}\right|^2,
\end{align}
where $R$ and ${\cal G}$ are defined as
\begin{align}
R^{-1}
=|\alpha_3|\left(g_1'(u_+-U_6|\tau)-g_1'(u_+-U_3|\tau)\right)\,,
\quad{\cal G}={2\pi i\over|\alpha_3|}\eta(\tau)^2\,,
\end{align}
with $g_1'(\nu|\tau)=\partial_{\nu}^2[\log\vartheta_1(\nu|\tau)]$ and
$\eta(\tau)=e^{\pi i\tau\over 12}
\prod_{n=1}^{\infty}(1-e^{2\pi in\tau})$.
The contribution from the matter part is
\begin{align}
&\left\langle :e^{ik_3X(U_3,\bar{U}_3)}:
:e^{ik_6X(U_6,\bar{U}_6)}:\alpha_1\alpha_2\,{\cal C}\right\rangle_\tau
=(2\pi {\rm Im}\tau)\int d\alpha_1
\delta(T_2-T_1)\nonumber\\
&\quad\times\delta^d(k_3+k_6)(2{\rm Im}\tau)^{-{d\over 2}}
e^{-{\pi k_3^2\over 2{\rm Im}\tau}
\bigl((U_3-U_6)-(\bar{U}_3-\bar{U}_6)\bigr)^2}
\left|\eta(\tau)\right|^{-2d}
\left|{\vartheta}_1(U_3-U_6|\tau)
\over\vartheta'_1(0|\tau)\right|^{-2k_3^2}\alpha_1\alpha_2\,{\cal C}\,.
\end{align}
The derivation of the first line from the zero mode sector of the
matter part is technical.
See \cite{AKT3} for more details.
Note that $\delta(T_2-T_1)=\delta(\rho(u+\tau)-\rho(u))$ implies 
that the correlator gives a nonzero result only when
\begin{align}
U_6-U_3=\frac{\alpha_1}{|\alpha_3|}\tau\,,
\end{align}
is satisfied.

Combining the ghost, matter contribution and the conformal factor, we
find ($d=26$)
\begin{align}
&\Big\langle\alpha_1\alpha_2
b_{T_1}\bar{b}_{T_1}b_{T_2}\bar{b}_{T_2}
V(k_3;U_3,\bar{U}_3)V(k_6;U_6,\bar{U}_6)\,{\cal C}\Big\rangle_\tau
\nonumber\\
&\sim(2\pi)^d\delta^d(k_3+k_6)\int d\alpha_1
\delta(T_2-T_1)2^{-28}\pi^{-13}
\frac{\alpha_1\alpha_2}{\alpha_3^4}
\biggl[{T\over|\alpha_3|}
\biggl(\log{T\over|\alpha_3|}\biggr)^2\,\biggr]^{-6}\,,
\label{alphap+}
\end{align}
for $T=T_1=T_2\to +0$.
Therefore we find $|B(3,6)\rangle$ is proportional to the reflector
with the singular coefficient as expected from \eqref{twosigma}:
\begin{align}
|B(3,6)\rangle\sim
2^{-26}\pi^{-12}
\biggl[{T\over|\alpha_{123}|}
\biggl(\log{T\over |\alpha_{123}|}\biggr)^2\,\biggr]^{-6}
|R(3,6)\rangle\,,\label{loopresult}
\end{align}
if we compare the result of LCSFT \eqref{eq:det_eval} and that of the 
$\alpha=p^+$ HIKKO string field theory \eqref{alphap+}.

\section{Discussion}
In this paper, we investigated the correspondence between the twist
field $\sigma(z,\bar z)$ and the three-string interaction vertex
$|V\rangle_{123}$ in LCSFT.
We evaluated two diagrams corresponding to the OPE \eqref{twosigma}
and found that both of them, \eqref{treeresult} and
\eqref{loopresult}, showed the same behavior including the log factor
as expected from the calculation of the twist field.

We would like to list several further directions.
\begin{itemize}
\item
Due to some technical difficulties, our computation of the Neumann
matrices is not completely satisfactory.
First of all, we only perform the numerical analysis for \eqref{CC/c}
instead of proving it algebraically.
Secondly, to evaluate $K_T$ we have to detour to the CFT techniques and
the $\alpha=p^+$ HIKKO string field theory.
We hope we will have more direct computation tools in the future.
\item
One of our original motivations comes from construction of LCSFT.
After relating the twist field $\sigma(z,\bar z)$ with the
three-string vertex $|V\rangle_{123}$ carefully in this paper, we
would like to see how matrix string theory can help in the
construction of LCSFT.
As explained in the introduction, it is difficult to proceed to
construction of higher order contact terms in LCSFT.
We would like to see whether we can construct higher order terms
explicitly with the help of MST.
Our realization of the twist field via the three-string interaction
vertex has been considered in the bosonic string theory in this
paper.
The first step should be to generalize our computation to the
superstring case.
\item
In the fermionic sector, it is a popular fact that the spin fields can
be realized by the fundamental free bosons.
It is interesting to see whether the free field realization has
anything to do with our realization via the three-string interaction
vertex in LCSFT.
Though there is no simple free field realization for the twist field,
we can construct one for its open string cousin, the boundary changing 
operator.
Since the boundary changing operator changes the boundary conditions
between the Neumann type and the Dirichlet type, or in other words,
changes the signs of the anti-holomorphic part, it can be regarded
as the twist field in the open string sector.
We present the free field realization of the boundary changing
operator in appendix B by applying a result of \cite{KT}.
Hopefully, we can relate the free field realization with our current
realization via the three-string vertex in the future.
\end{itemize}

\section*{Acknowledgments}
We would like to thank A.~Hashimoto, P.~Ho, Y.~Kazama, E.~Watanabe and
especially H.~Hata for valuable discussions and comments.
The work of S.\,T. was supported by the National Center for
Theoretical Sciences (North), (NSC 94-2119-M-002-001).
S.\,M. would like to thank University of Tokyo (Komaba) and KEK for
hospitality where part of this work was done. 
Part of this work was done in 2005 Taipei Summer Institute on Strings,
Particles and Fields, Summer Institute String Theory 2005 at Sapporo
and YITP workshop (YITP-W-05-21).
We are grateful to the organizers of these workshops.

\appendix
\section{Neumann coefficients}
In this appendix, we would like to briefly review the Neumann
coefficients.
The Neumann coefficient matrix are constructed from the overlapping
condition
\begin{align}
&\Bigl(P^{(1)}(\sigma)\Theta(-\pi\alpha_1<\sigma<\pi\alpha_1)
+P^{(2)}(\sigma-\pi\alpha_1)
\Theta(\pi\alpha_1<\sigma)
+P^{(2)}(\sigma+\pi\alpha_1)
\Theta(\sigma<-\pi\alpha_1)\nonumber\\
&\qquad
+P^{(3)}(\pi(\alpha_1+\alpha_2)-\sigma)\Theta(0<\sigma)
+P^{(3)}(-\pi(\alpha_1+\alpha_2)-\sigma)
\Theta(\sigma<0)\Bigr)|V\rangle_{123}=0\,,
\end{align}
of the momentum function of each string
\begin{align}
P^{(r)}(\sigma)=\frac{1}{2\pi|\alpha_r|}\biggl[p^{(r)}
+\sum_{n=1}^\infty\sqrt{2n}
\biggl(p^{(r){\rm c}}_n\cos\frac{n\sigma}{|\alpha_r|}
+p^{(r){\rm s}}_n\sin\frac{n\sigma}{|\alpha_r|}\biggr)\biggr]\,,
\label{P}
\end{align}
which states that in the string interaction process the momentum is
conserved along the string worldsheet.
$\Theta({\rm inequality})$ denotes the step function, which takes
value $1$ if the inequality holds and otherwise $0$.

Our next task is to rewrite the overlapping condition in terms of the
mode expansion.
If we normalize the momentum $\boldsymbol{\pi}$ so that the
trigonometric functions have the unit norm:
\begin{align}
\boldsymbol{\pi}^{(i){\rm c}}=\frac{1}{2\pi|\alpha_i|}
\begin{pmatrix}p^{(i)}\\
\sqrt{C}\boldsymbol{p}^{(i){\rm c}}\end{pmatrix},
\quad\boldsymbol{\pi}^{(i){\rm s}}=\frac{1}{2\pi|\alpha_i|}
\sqrt{C}\boldsymbol{p}^{(i){\rm s}},
\end{align}
the overlapping condition can be recast into the following form
\begin{align}
\boldsymbol{\pi}^{(3){\rm c}}=
\begin{pmatrix}U_1&U_2\end{pmatrix}
\begin{pmatrix}\sqrt{-\alpha_1/\alpha_3}
\boldsymbol{\pi}^{(1){\rm c}}\\
\sqrt{-\alpha_2/\alpha_3}
\boldsymbol{\pi}^{(2){\rm c}}\end{pmatrix},\quad
\boldsymbol{\pi}^{(3){\rm s}}=
\begin{pmatrix}V_1&V_2\end{pmatrix}
\begin{pmatrix}\sqrt{-\alpha_1/\alpha_3}
\boldsymbol{\pi}^{(1){\rm s}}\\
\sqrt{-\alpha_2/\alpha_3}
\boldsymbol{\pi}^{(2){\rm s}}\end{pmatrix},
\end{align}
where $U_1$, $U_2$, $V_1$, $V_2$ are defined as
\begin{align}
&U_1=\begin{pmatrix}
-\sqrt{-\alpha_1/\alpha_3}&\boldsymbol{0}^{{\rm T}}\\
\displaystyle{\sqrt{-\frac{\alpha_1}{\alpha_3}}}\alpha_2
\displaystyle{\sqrt{\frac{C}{2}}}\boldsymbol{B}&
-\displaystyle{\sqrt{-\frac{\alpha_1}{\alpha_3}}}\sqrt{C}A^{(1)}
\displaystyle{\frac{1}{\sqrt{C}}}
\end{pmatrix},\nonumber\\
&U_2=\begin{pmatrix}
-\sqrt{-\alpha_2/\alpha_3}&\boldsymbol{0}^{{\rm T}}\\
-\displaystyle{\sqrt{-\frac{\alpha_2}{\alpha_3}}}\alpha_1
\sqrt{\frac{C}{2}}\boldsymbol{B}&
-\displaystyle{\sqrt{-\frac{\alpha_2}{\alpha_3}}}
\sqrt{C}A^{(2)}\frac{1}{\sqrt{C}}
\end{pmatrix},\\
&V_1=\sqrt{-\frac{\alpha_3}{\alpha_1}}
\frac{1}{\sqrt{C}}A^{(1)}\sqrt{C}\,,\quad
V_2=\sqrt{-\frac{\alpha_3}{\alpha_2}}
\frac{1}{\sqrt{C}}A^{(2)}\sqrt{C}\,,\nonumber
\end{align}
with $A^{(1)}$, $A^{(2)}$, $\boldsymbol{B}$ and $C$ being
\begin{align}
(A^{(1)})_{mn}&=\sqrt{\frac{n}{m}}\frac{(-1)^m}{\pi\alpha_1}
\int_0^{\pi\alpha_1}\!
2\cos\frac{n\sigma}{\alpha_1}\cos\frac{m\sigma}{\alpha_3}d\sigma
=\sqrt{\frac{m}{n}}\frac{(-1)^m}{\pi\alpha_3}
\int_0^{\pi\alpha_1}\!
2\sin\frac{n\sigma}{\alpha_1}\sin\frac{m\sigma}{\alpha_3}d\sigma\,,\\
(A^{(2)})_{mn}&=\sqrt{\frac{n}{m}}\frac{(-1)^m}{\pi\alpha_2}
\int_{\pi\alpha_1}^{\pi(\alpha_1+\alpha_2)}\!
2\cos\frac{n(\sigma-\pi\alpha_1)}{\alpha_2}
\cos\frac{m\sigma}{\alpha_3}d\sigma\nonumber\\
&=\sqrt{\frac{m}{n}}\frac{(-1)^m}{\pi\alpha_3}
\int_{\pi\alpha_1}^{\pi(\alpha_1+\alpha_2)}\!
2\sin\frac{n(\sigma-\pi\alpha_1)}{\alpha_2}
\sin\frac{m\sigma}{\alpha_3}d\sigma\,,\\
(\boldsymbol{B})_m&=\frac{2(-1)^{m+1}}{\sqrt{m}\pi\alpha_1\alpha_2}
\int_0^{\pi\alpha_1}\!\cos\frac{m\sigma}{\alpha_3}d\sigma
=\frac{2(-1)^{m}}{\sqrt{m}\pi\alpha_1\alpha_2}
\int_{\pi\alpha_1}^{\pi(\alpha_1+\alpha_2)}\!
\cos\frac{m\sigma}{\alpha_3}d\sigma\,,\\
(C)_{mn}&=m\delta_{mn}\,.
\end{align}

Since the overlapping condition relates the incoming string momentum
with the outgoing string momentum, it does not drop any information.
Therefore, the transformation matrices
$\begin{pmatrix}U_1&U_2\end{pmatrix}$ and
$\begin{pmatrix}V_1&V_2\end{pmatrix}$ should be unitary \cite{GMP3}.
By requiring the unitarity of these matrices, we have ($r,s=1,2$)
\begin{align}
&-\frac{\alpha_r}{\alpha_3}
A^{(r){\rm T}}CA^{(s)}
=\delta_{rs}C\,,\quad
A^{(r){\rm T}}C\boldsymbol{B}=0\,,\quad
\frac{1}{2}\alpha_1\alpha_2
\boldsymbol{B}^{{\rm T}}C\boldsymbol{B}=1\,,\quad
-\frac{\alpha_3}{\alpha_r}
A^{(r){\rm T}}\frac{1}{C}A^{(s)}=\delta_{rs}\frac{1}{C}\,,
\nonumber\\
&\qquad\qquad\sum_{t=1}^3\alpha_tA^{(t)}\frac{1}{C}A^{(t){\rm T}}
=\frac{1}{2}\alpha_1\alpha_2\alpha_3
\boldsymbol{B}\boldsymbol{B}^{{\rm T}}\,,\quad
\sum_{r=1}^3\frac{1}{\alpha_r}A^{(r)}CA^{(r){\rm T}}=0\,,
\end{align}
if we define $(A^{(3)})_{mn}=\delta_{mn}$ in addition.
In fact, these identities are proved in \cite{GS}.

As in \cite{GS}, the three-string interaction vertex can be
constructed by matching the momentum eigenstates.
After the Gaussian integration, we find the result is given by
\eqref{3string} with the Neumann coefficient matrices given as
\begin{align}
N^{rs}=\delta^{rs}-2A^{(r)\rm T}\Gamma^{-1}A^{(s)}\,,\quad
\boldsymbol{N}^r=-A^{(r)\rm T}\Gamma^{-1}\boldsymbol{B}\,,\quad
\Gamma=1+A^{(1)}A^{(1)\rm T}+A^{(2)}A^{(2)\rm T}\,,
\end{align}
With this formal expression of the Neumann coefficients, we can prove
all the formulas we need in this paper \eqref{NN}, as well as \cite{Y}
\begin{align}
\sum_{t=1}^3N^{r,t}N^{t,s}=\delta_{r,s}\,,\quad
\sum_{t=1}^3N^{r,t}\boldsymbol{N}^t=-\boldsymbol{N}^r\,,\quad
\sum_{t=1}^3\boldsymbol{N}^{t{\rm T}}\boldsymbol{N}^t
={2\tau_0\over \alpha_1\alpha_2\alpha_3}\,.
\end{align}

\section{Free field realization of boundary changing operators}
In this appendix we would like to present a free field realization of
the boundary changing operators.
In \cite{CKLM} a certain class of boundary deformations
\begin{align}
S_{\rm int}=-\frac{1}{2}\int d\theta
\biggl(g\exp\frac{iX(\theta)}{\sqrt{2}}+
\bar g\exp\frac{-iX(\theta)}{\sqrt{2}}\biggr)\,,
\end{align}
was solved exactly by the boundary state
\begin{align}
\langle B|=\langle N|\exp\bigl(-i\pi(g_rJ^+_0+\bar g_rJ^-_0)\bigr)\,,
\end{align}
when the target space is compactified at the self-dual radius.
Here $g_r$ is the renormalized coupling constant, which equals to $0$
when the boundary interaction satisfies the Neumann boundary
condition, $g=0$, and to $1/2$ when it satisfies the Dirichlet
boundary condition, $g=\infty$.
In \cite{KT} the relation between the bare coupling constant $g$ and
the renormalized one $g_r$ was worked out explicitly and the coupling
constant $g$ is further generalized into an external source
$g(\theta)$ depending on the boundary coordinate $\theta$.
The result is
\begin{align}
&\langle B|=\langle N|\exp\left[\int
d\theta\left(\frac{1}{2}g_r(g(\theta),\bar g(\theta))
e^{i\sqrt{2}X_L(\theta)}
+\frac{1}{2}\bar g_r(g(\theta),\bar g(\theta))
e^{-i\sqrt{2}X_L(\theta)}
\right)\right]\,,\\
&\qquad\qquad g_r(g(\theta),\bar g(\theta))
=\frac{2}{\pi}\frac{g(\theta)}{|g(\theta)|}
\arctan\Bigl[\tanh\Bigl(\frac{\pi}{2}|g(\theta)|\Bigr)\Bigr]\,.
\end{align}

If we choose $g(\theta)$ to be
\begin{align}
g(\theta)=g\Theta(\phi_1<\theta<\phi_2)\,,
\end{align}
and take the limit $g\to\infty$ finally, we can realize a situation
with one part of the boundary satisfying the Neumann boundary
condition while the other satisfying the Dirichlet boundary
condition.
This boundary is the same as that with two boundary changing operators
inserted.
Hence we can consider the realization of the boundary changing
operators using the result of \cite{KT}.
With this choice of $g(\theta)$ we have
\begin{align}
g_r(g(\theta),\bar g(\theta))=
\frac{1}{2}\Theta(\phi_1<\theta<\phi_2)\,.
\end{align}
In terms of the mode expansion
$J^a(z)=\sum_{m=-\infty}^\infty J^a_m/z^{m+1}$, our result is
expressed by
\begin{align}
\langle B|=\langle N|\exp\sum_{n=-\infty}^{\infty}
\frac{(w^{-n}_1-w^{-n}_2)}{2n}J_n^1\,,
\end{align}
with $w_i=e^{-i\phi_i}$.
Here we have used the Fourier expansion of the step function,
\begin{align}
\Theta(\phi_1<\theta<\phi_2)=
\frac{2}{\pi}\sum_{n=-\infty}^{\infty}
\frac{ie^{in\theta}(e^{-in\phi_2}-e^{-in\phi_1})}{4n}\,.
\end{align}
Note that we have used the bookkeeping notation for the zero mode
$n=0$.
More correctly, the zero mode should be spelled out as
$(\phi_2-\phi_1)/4$.

Therefore, a naive candidate for the boundary changing operators is
\begin{align}
\sigma_\pm(w)\simeq\exp\pm\Biggl(\sum_{m\neq 0}\frac{w^{-m}}{2m}J_m^1
-\frac{\log w}{2}J_0^1\Biggr)\,.
\label{naive}
\end{align}
The sign $\pm$ is chosen depending whether the boundary operator
changes the Neumann type into the Dirichlet type or vice versa.
In the expression of \eqref{naive} we are not careful about the zero
mode and the normal ordering.
This ambiguity can be fixed by requiring that the boundary changing
operators are the primary fields with dimension (1/16,0).
Since we have
\begin{align}
\sum_{m\neq 0}\frac{w^{-m}}{2m}J_m^3
-\frac{\log w}{2}J_0^3-\frac{i}{2\sqrt{2}}x_L
=-\frac{iX_L(w)}{2\sqrt{2}}\,,
\end{align}
if we replace the direction $1$ by the direction $3$ in the exponent
of \eqref{naive}, our boundary changing operators should be written as
\begin{align}
\sigma_\pm(w)=e^{-i\pi J^2_0/2}
:\exp\pm\left(-\frac{iX_L(w)}{2\sqrt{2}}\right):
e^{i\pi J^2_0/2}\,,
\label{freefield}
\end{align}
which are $(1/16,0)$ primary fields.

It is difficult to proceed to simplify the expression
\eqref{freefield} of the boundary changing operators but we can check
the OPEs
\begin{align}
\sigma_+(z)\cdot\sigma_-(0)\sim\frac{1}{z^{1/8}}\,,\quad
\partial X_L(z)\cdot\sigma_\pm(0)\sim\frac{1}{\sqrt{z}}\tau_\pm(0)\,.
\label{twoope}
\end{align}
The latter one is shown by transforming the operator $\partial X_L(z)$
by the SU(2) rotation, instead of rotating the boundary changing
operators.
Since under the $J^2_0$ rotation $J^3(z)=i\partial X_L(z)/\sqrt{2}$ is
transformed into $J^1(z)=:\cos\sqrt{2}X_L(z):$, the latter one of
\eqref{twoope} is easily found by noting
\begin{align}
:e^{i\sqrt{2}X_L(z)}:\,
\cdot:e^{-\frac{i}{2\sqrt{2}}X_L(0)}:
\sim\frac{1}{\sqrt{z}}
:e^{\frac{3i}{2\sqrt{2}}X_L(0)}:\,.
\end{align}
Here we have used $X_L(z)\cdot X_L(0)\sim -\ln z$.

\end{document}